\documentclass [preprint,preprintnumbers,amsmath,amssymb] {revtex4}
\usepackage[dvipdfmx]{graphicx}
\usepackage{bm}

\usepackage{amsmath}	
\begin{document}
\title{Parity reversion of $^{12}_{\Lambda}$Be}
\author{H. Homma$^1$, M. Isaka$^1$ and M. Kimura$^2$}
\affiliation{$^1$Department of Cosmosciences, Graduate School of Science, Hokkaido University,
Sapporo 060-0810, Japan\\
$^2$Creative Research Institution (CRIS), Hokkaido University, Sapporo 001-0021,
Japan}

\begin{abstract}
The spectrum of $^{12}_\Lambda$Be is studied by an extended version of antisymmetrized
 molecular dynamics for hypernuclei. The result predicts the positive-parity ground state of
 $^{12}_\Lambda$Be that is  reverted to the  normal one by the impurity effect of $\Lambda$
 particle. The reversion of the parity is due  to the difference of $\Lambda$ binding energy
 in the positive- and negative-parity states  that originates in the difference of  $\alpha$
 clustering and deformation.  
\end{abstract}

\maketitle

Despite of their short lifetime, hypernuclei have been a subject of particular interest in nuclear
physics.  They provide an almost unique opportunity to investigate underlying baryon-baryon
interactions. Especially the knowledge of the interaction between $\Lambda$ and nucleons is
greatly increased in these decades \cite{Nag75,Yam83,Mil85,Reu94,Rij06,Fuj07}. This development
strongly promotes the physics of hypernuclear many-body problems. 
 A peculiar interest of hypernucler many-body
problems is the dynamical features $\Lambda$ hypernuclei manifest by the addition of
$\Lambda$ particle such as the stabilization of the system \cite{Mod90,Ohk95}, modifications of
sizes \cite{Hiy99,Tan01}, deformation  and clustering
\cite{Yam84,Sak87,Win08,Sch10,Isa11-1,Isa11-2}. Thus, hypernuclear many-body physics can be
regarded as an impurity physics. It offers a means to investigate dynamical responses of
nuclei to the addition of hyperons or a means to investigate nuclear structure by using
hyperons as probe. As the forthcoming experiments will expand domain of hypernuclear physics
to neutron-rich or heavier system, many exotic phenomena due to the impurity effect of
$\Lambda$ will be uncovered. 

Single $\Lambda$ hypernuclei of neutron-rich Be isotopes are one of such systems of interest 
and to be experimentally accessible. More specifically, $^{12}_\Lambda$Be is of particular
interest, since the core nucleus $^{11}$Be is known to have quite exotic structure. It has
two bound states with positive- and negative-parity. The ground state is positive parity
and the negative-parity state is located at 320 keV above the ground state
\cite{Wil59,Tal60,Ajz75}. Since the order of these two states  contradicts to the ordinary nuclear
shell ordering and the neutron shell  gap of $N=8$ is collapsed, it is called ``parity
inversion''  and ``breakdown of $N=8$ magic number''. Our question is how the ``parity
reversion'' will be affected and modified in $^{12}_\Lambda$Be by the impurity effect of
$\Lambda$. This letter reports that the parity inverted in  $^{11}$Be will be reverted in
$^{12}_\Lambda$Be by adding  $\Lambda$ particle. This study is based on the theoretical
framework of antisymmetrized molecular dynamics (AMD). AMD has been applied to investigate
the exotic phenomena in neutron-rich nuclei and has successfully described them such as the
breakdown of $N=8$ and  $N=20$  magic  numbers
\cite{Dot97,Kan02,Kan03-2,Oer06,Kim04,Kim07}. In this study, we use an extended
version of AMD for hypernuclei (HyperAMD) to investigate $^{12}_\Lambda$Be. HyperAMD has
already been applied to $p$-$sd$ shell hypernuclei \cite{Isa11-1,Isa11-2} and the reader is
directed to them for its detailed formulation.   The Hamiltonian used in this study is given as, 
\begin{align}
H &= H_{N} + H_{\Lambda}  - T_g,\\
H_{N} &= T_{N} + V_{NN} + V_{C}, \quad H_{\Lambda}=T_{\Lambda} + V_{\Lambda N},
\end{align}
where $T_{N},  T_{\Lambda}$ and $T_{g}$ are the kinetic energies of the
nucleons, $\Lambda$ particle and the center-of-mass motion. The Gogny D1S
\cite{Gog80} is used as an effective nucleon-nucleon interaction $V_{NN}$, and the
Coulomb interaction $V_{C}$ is approximated by the sum of seven Gaussians. 
To see the dependence on  $\Lambda N$ interaction, a couple of $\Lambda N$ effective
interactions $V_{\Lambda N}$ are examined. We have used $YN$ G-matrix interactions YNG-ND
 and YNG-NF \cite{Yam83} which are respectively derived from the realistic
one-boson-exchange potentials of the Nijmegen model-D and model-F \cite{Nag75}. We
also used a modified version of YNG-NF (Improved-NF) suggested by Hiyama   
{\it et al.} \cite{Hiy97}. These $\Lambda N$ interactions have the dependence on the nucleon
Fermi momentum $k_F$ and the value of $k_F=0.973$ fm$^{-1}$  is applied, that is common to
the Ref. \cite{Hiy97}. 

The variational wave function of HyperAMD is the eigenstate of the parity. The intrinsic
wave function $\Psi_{int}$ is represented by the direct product of the $\Lambda$ single
particle wave function $\varphi_\Lambda$ and the wave function of the $A$ nucleons $\Psi_N$
which is a Slater determinant of the nucleon wave packets $\psi_i$,  
\begin{align}
\Psi^{\pm} &= P^{\pm} \Psi_{int},\quad \Psi_{int} = \Psi_{N} \otimes \varphi_{Y},\\
\Psi_{N} &=  \frac{1}{\sqrt{A!}} \det \{ \psi_{i} \left( \bm r_j \right) \},
\end{align}
where $P^\pm$ is the parity projector. 
The nucleon single particle wave packet is represented by a Gaussian,
\begin{align}
\psi_{i} \left( \bm{r} \right) &= \phi_{i} \left(\bm{r} \right) \chi_{i} \tau_{i}, \\
\phi_{i}\left( \bm{r} \right) &= \prod_{\sigma = x,y,z} \left( \frac{2\nu_{\sigma}}{\pi} \right)^{1/4}
\exp \{ - \nu_{\sigma} \left( r - Z_{i} \right)^{2}_{\sigma} \}, \\
\chi_{i} &= a_{i}\chi_{\uparrow} + b_{i}\chi_{\downarrow}, \quad
\tau_{i} = \text{p or n}. 
\end{align}
The $\Lambda$ single particle wave function is represented by a sum of Gaussians, 
\begin{align}
\varphi_{\Lambda} \left( \bf{r} \right) &= \sum_{m=1}^{M} c_{m} \varphi_{m} \left(\bf{r} \right)
, \quad \varphi_{m} \left( \bf{r} \right) = \phi_{m} \left( \bf{r} \right) \chi_{m}, \\
\phi_{m} \left( \bf{r} \right) &= \prod_{\sigma = x,y,z} \left( \frac{2\nu_{\sigma}}{\pi} \right)^{1/4}
\exp \{ - \nu_{\sigma} \left( r - \zeta_m  \right)^{2}_{\sigma} \}, \\
\chi_{m} &= \alpha_{m}\chi_{\uparrow} + \beta_{m}\chi_{\downarrow},
\end{align}
where the number of Gaussians $M$ is chosen large enough to achieve the energy
convergence. The centroids of Gaussian wave packets $\bm Z_i$ and $\bm \zeta_m$, the width
of Gaussian $\nu_\sigma$, the coefficients $c_m$ and spin directions $a_i, b_i, \alpha_m$,
$\beta_m$ are the  variational parameters. They are so determined to minimize the total 
energy under the constraint on the matter quadrupole deformation parameter $\beta$
\cite{Dot97}.  The constraint is imposed on the value of $\beta$ from 0 to 1.2 with the interval of 0.025.

After the variation, we project out the eigenstate of the total angular momentum $J$ for
each value of $\beta$, 
 \begin{align}
 \Psi^{J\pm}_{MK}(\beta) &=
\frac{2J+1}{8\pi^2}\int d\Omega D^{J*}_{MK}(\Omega)R(\Omega) \Psi^\pm(\beta).
 \end{align}
The integrals over three Euler angles $\Omega$ are performed numerically. The calculation is
completed by the generator coordinate method (GCM) \cite{Hil53}. The wave functions that have
different values of $K$ and $\beta$ are superposed,
\begin{align}
 \Psi_n^{J\pm}&=\sum_p\sum_{K=-J}^{J} c_{npK} \Psi^{J\pm}_{MK}(\beta_p). 
 \label{eq:GCM}
\end{align}
The coefficients $c_{npK}$ are determined by solving Griffin-Hill-Wheeler equation. 

\begin{figure*}[ht]
  \centering
  \includegraphics[width=0.75\hsize]{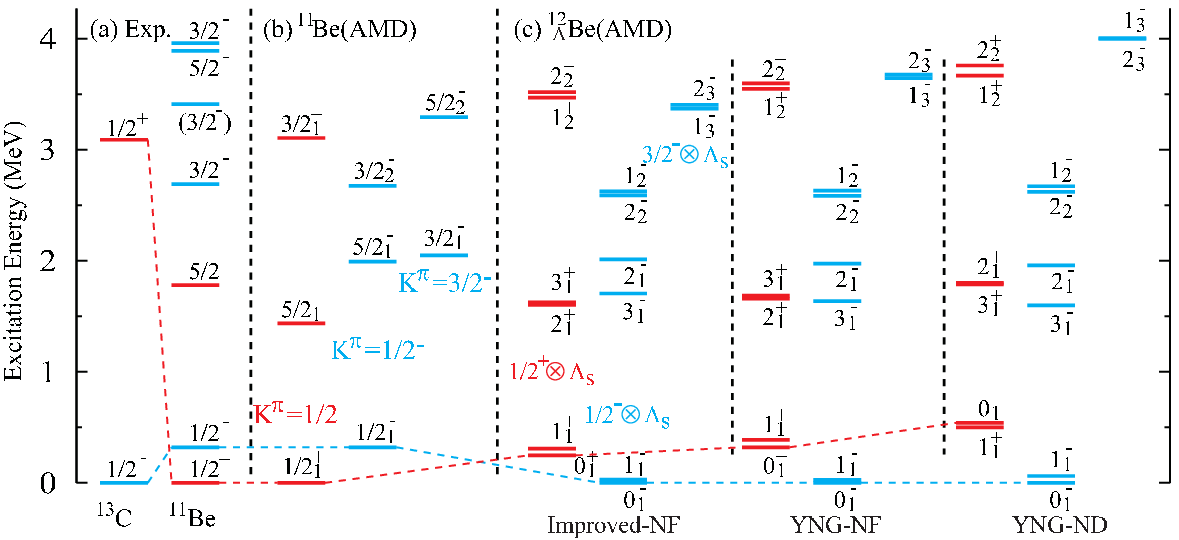}
  \caption{(color online) (a) Observed spectra of $^{13}$C and $^{11}$Be. (b) Spectrum of
 $^{11}$Be calculated by AMD. (c) Spectra of $^{12}_\Lambda$Be calculated by HyperAMD with 
 three different $\Lambda N$ interactions.}
  \label{fig:LevelScheme}
\end{figure*}
\begin{figure}
  \centering
  \includegraphics[width=\hsize]{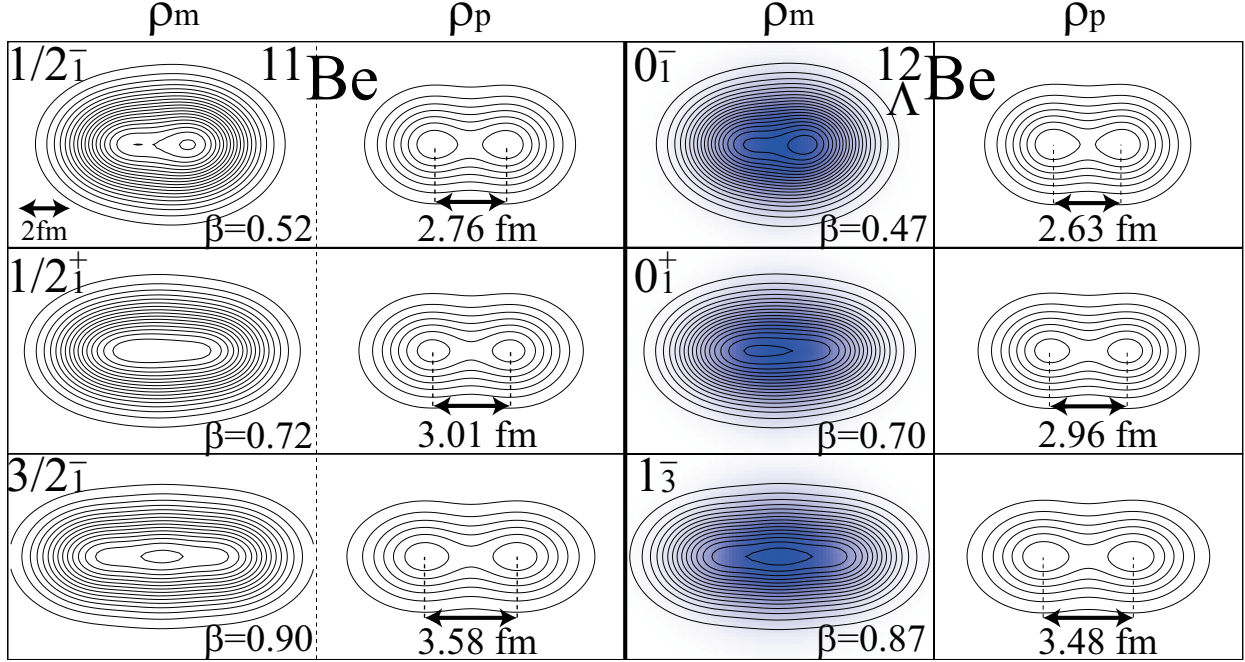}
  \caption{(color online) Matter ($\rho_m$) and  proton ($\rho_p$) density distributions of
 the each  band head  states of $^{11}$Be and $^{12}_{\Lambda}$Be.   Color plots show the
 density  distribution  of $\Lambda$ particle.}  
  \label{fig:Density}
\end{figure}%
\begin{figure}
  \centering
  \includegraphics[width=\hsize]{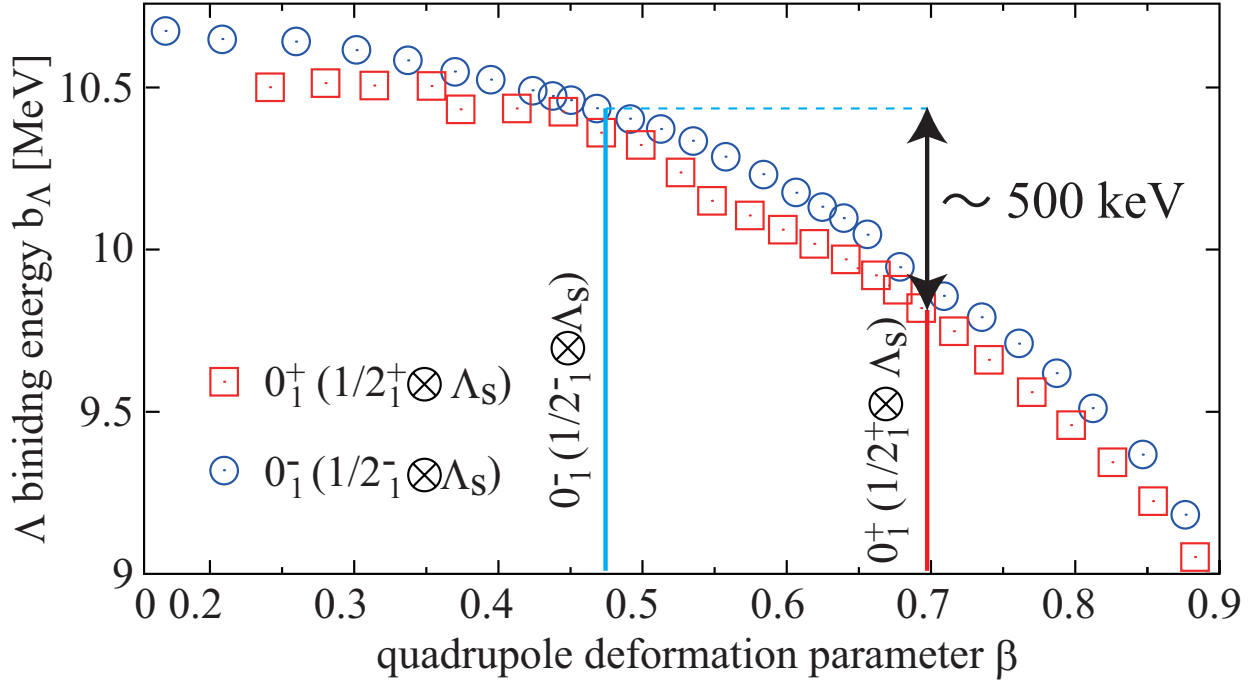}
  \caption{(color online) The $\Lambda$ binding energy  of the $0^-_1$ and
 $0^+_1$ states as function of proton quadrupole deformation parameter$\beta$. Lines in the
 figure denote the deformation parameter of $0^\pm_1$ states given in Table. \ref{tab:tab}}
  \label{fig:BL}
\end{figure}%

\begin{table}[ht]
\caption{Calculated binding and excitation energies $B$ and $E_x$ [MeV], matter quadrupole
 deformation $\beta$ and root-mean-square radii $r_{rms}$ [fm] for band-head states of
 $^{11}$Be and $^{12}_\Lambda$Be. Numbers in parenthesis shows the observed data
 \cite{Oza01}. The $\Lambda$ binding energy  $B_\Lambda$ [MeV],   the of $\Lambda$ kinetic
 energy  $T_\Lambda$ [MeV] and  $\Lambda N$ potential energy $V_{\Lambda N}$ [MeV] are also
 shown  for  $^{12}_\Lambda$Be.} \label{tab:tab}  
\begin{center}
  \begin{tabular}{ccccccccc} \hline \hline
   &$J^\pi$ & $B$ & $E_x$ & $\beta$ & $r_{rms}$&
   $B_\Lambda$ &$T_\Lambda$ &$V_{\Lambda N}$\\
   \hline
   &   $1/2^-_1$ & 64.45 & 0.32 & 0.52 & 2.53& & &\\
   &   & (65.16) & (0.32) & &  & & &\\
   $^{11}$Be &  $1/2^+_1$ & 64.77 & 0 & 0.72 & 2.69 & & &\\
   &   & (65.48) &(0)  & & (2.73) & & &\\
   &   $3/2^-_1$ & 62.72 & 2.05 & 0.90 & 2.98 & & &\\
   \hline
   &   $0^-_1$ & 74.69 & 0 & 0.47 & 2.51& 10.24 &6.71 & -16.93\\
   $^{12}_\Lambda$Be &   $0^+_1$ & 74.44 & 0.25 & 0.70 & 2.67& 9.67 &6.68&-16.42\\
   &   $1^-_3$ & 71.32 & 3.37 & 0.87 & 2.94 & 8.60 & 6.36&-15.08\\
   \hline\hline
\end{tabular}
\end{center}
\end{table}%


Before the discussion on $^{12}_\Lambda$Be, it is helpful to overview the structure of
$^{11}$Be.
Figure \ref{fig:LevelScheme} (a) shows the observed spectra of $N=7$ isotones, $^{13}$C and 
$^{11}$Be. The large shell gap between $p_{1/2}$ and
$sd$-shell  in $^{13}$C is collapsed in $^{11}$Be. Namely, the ground state is
positive parity and the order of 
$p_{3/2}$ and $sd$-shell looks inverted in $^{11}$Be, that is called ``parity inversion'' \cite{Wil59,Tal60}.
Using the original parameter set of the Gogny D1S, our calculation successfully reproduces
the spin-parity of the ground state with the binding energy of 65.32 MeV and the first
excited state  $1/2^-_1$ is located at 540 keV, while the observed
values  are 65.48 MeV and 320 keV, respectively \cite{Ajz75}. For more quantitative discussion
of $^{12}_\Lambda$Be, we have weakened the spin-orbit interaction of Gogny D1S by 5\% to 
reproduce the observed $1/2^-_1$ excitation energy exactly.  By this modification, the binding
energy of $^{11}$Be is calculated as 64.77 MeV and the resulting spectrum is shown in
Fig. \ref{fig:LevelScheme} (b). Here, the excited unbound states are calculated within the bound
state approximation.

It is known that the low-lying states of Be isotopes have 2$\alpha$ cluster core and
valence neutrons occupying the molecular-orbits around the core which are so-called $\pi$ and
$\sigma$ orbits \cite{Oer06}. The formation of the 2$\alpha$ cluster core  in each state is
confirmed in the proton density shown in Fig. \ref{fig:Density}.  The ground state is a member of the
$K^\pi=1/2^+$ band in which two of three valence neutrons occupy $\pi$-orbit and the last
valence neutron occupies $\sigma$-orbit. In terms of the spherical shell model,  a neutron
is promoted into $sd$-shell across the $N=8$ shell gap (breakdown of magic number
$N=8$). The first excited state is negative parity and belongs to the $K^\pi=1/2^-$
band. All valence neutrons occupy $\pi$-orbit or $p$-shell, that corresponds to the normal
shell order. As we can see in Fig. \ref{fig:Density} and Table \ref{tab:tab}, the ground
state has more pronounced 2$\alpha$ clustering and larger quadrupole deformation $\beta$
than the first excited state. Here, deformation $\beta$ of each state
is defined as that of the basis wave function $\Psi^{J\pm}_{MK}(\beta)$ which has the
maximum overlap with the GCM wave function (Eq. \ref{eq:GCM}). 
The $3/2^-_2$ state (band head of $K^\pi=3/2^-$) that has two valence neutrons in
$\sigma$-orbit is most deformed amongst the band-head states. Including the $3/2^-_2$ state,
the assignment of the unbound excited states is still under discussion, and we
devolve further discussions  to the Refs. \cite{Fre10,Pet11,For11}. We just remark here that
the ground state is more deformed than the first excited state and has a neutron in $sd$-shell.

The spectra of $^{12}_\Lambda$Be obtained with three different $\Lambda N$ interactions are
shown in Fig. \ref{fig:LevelScheme} (c). All states shown in the figure have a $\Lambda$ in
$s$-orbit and are classified into three bands. They are generated by the coupling of $K^\pi=1/2^+, 1/2^-$
and $3/2^-$ bands of $^{11}$Be with $\Lambda$ in $s$-orbit, and therefore, there are always
doublet states of $^{12}_\Lambda$Be for each corresponding state of $^{11}$Be. 
These bands are denoted as $K^\pi=1/2^+\otimes \Lambda_s, 1/2^-\otimes \Lambda_s$ and
$3/2^-\otimes \Lambda_s$. 

It is found that all of three $\Lambda N$ interactions predicts the negative-parity ground
state and give qualitatively same results. Therefore,  we focus on the Improved-NF result
for a while.  The ground doublet is the $0^-_1$ and $1^-_1$ states 
with the binding energies of 74.69 and 74.66 MeV, that have the configuration of
$^{11}$Be$(1/2^-_1)\otimes \Lambda_s$.  The first excited doublet is $0^+_1$ and $1^+_1$
states at 250 and 310 keV with the configuration of  $^{11}$Be$(1/2^+_1)\otimes
\Lambda_s$. Thus the ground state parity  is reverted in $^{12}_\Lambda$Be, as if  the
addition of $\Lambda$ has restored the $N=8$ shell gap.   This parity reversion of
$^{12}_\Lambda$Be is due to the difference of $\Lambda$ binding 
energy $B_\Lambda$ in the ground and first excited doublets. As shown in Table. \ref{tab:tab}, the
$0^-_1$ state has larger $B_\Lambda$ than the $0^+_1$ state by about 500 keV. Here
$B_\Lambda$ is defined as the difference of binding energies between $^{12}_\Lambda$Be and
corresponding $^{11}$Be states, 
\begin{align}
 B_\Lambda&= B(^{12}_\Lambda{\rm Be}(J^{\pi})) - B(^{11}{\rm Be}(J^{\prime\pi})) .
\end{align}
The change of the nuclear part $H_N$ by addition of $\Lambda$ is rather small. 
Therefore the difference of $B_\Lambda$ overwhelms the energy difference between the
$1/2^\pm_1$ states of $^{11}$Be, and the parity reversion is realized in
$^{12}_\Lambda$Be. The difference in $B_\Lambda$ mainly  comes from the difference in the
$\Lambda N$ potential $V_{\Lambda N}$ as shown in Table. \ref{tab:tab}, and it originates
in the difference of the quadrupole deformation. Figure \ref{fig:BL} shows the $\Lambda$ binding
energies in $0^\pm_1$ states as function of quadrupole deformation, that is defined as the
expectation value of $H_\Lambda$ by the angular-momentum projected wave functions for each
deformation parameter $\beta$,
\begin{align}
 b^\pm_\Lambda(\beta)&=-\langle \Psi^{0\pm}(\beta)|H_\Lambda|\Psi^{0\pm}(\beta)\rangle.
\end{align}
The $b_\Lambda$ of $0^+_1$ and $0^-_1$ states rapidly decrease as deformation becomes larger
and their behavior are quite similar to each other. As deformation becomes larger, the
overlap between the wave functions of $\Lambda$ and nucleons decreases to reduce $V_{\Lambda
N}$. We can confirm that  the $500$ keV
difference in $B_\Lambda$ originates in the different deformation of $0^\pm_1$ states from
Fig. \ref{fig:BL}.  This 
reduction of $b_\Lambda$ as function of 
quadrupole deformation is qualitatively common to other $p$-$sd$-shell nuclei discussed in
Ref. \cite{Isa11-1,Isa11-2}. Since the $3/2^-_1$ state is most deformed among the band head states of
$^{11}$Be, the $1^-_3$  state ($^{11}$Be$(3/2^-_1)\otimes\Lambda_s$) has smallest
$B_\Lambda$.  Therefore, its excitation energy is also shifted up compared to $^{11}$Be.   
The reduction of $B_\Lambda$ is common to other member states of $1/2^-\otimes \Lambda_s$
and $3/2^-\otimes \Lambda_s$ bands.

By the addition of $\Lambda$ particle, structure of core nucleus $^{11}$Be is slightly
modified. Due to the attraction of $\Lambda$ particle sitting at the center of the system,
the inter-cluster distance between 2$\alpha$ clusters is reduced (Fig. \ref{fig:Density}), and it leads to the
reduction of the deformation and radius (Table. \ref{tab:tab}). Dispite of the neutron-halo
structure of $^{11}$Be, this reduction if rather small compared to the obsebed case of
$^{7}_\Lambda$Li \cite{Tan01} and cannot be clearly seen in the
density distribution (Fig. \ref{fig:Density}).
This is due to the presence of valence  neutrons occupying molecular-orbits in 
$^{12}_\Lambda$Be. When the distance between 2$\alpha$ is reduced, the valence neutrons in
$\pi$ or $\sigma$-orbits loose their binding energies \cite{Oer06}. Therefore valence neutrons prevent drastic
reduction of 2$\alpha$ distance. Note that this explains why the change of the expectation
value of nuclear part is not large compared to the difference of $B_\Lambda$.   Other
interactions also predict the parity reversion of $^{12}_\Lambda$Be, and the mechanism is
common to the case of Improved-NF. Modification of nuclear structure is not large, but the
different deformation between the states leads to the difference in $B_\Lambda$ to revert
the parity of $^{12}_\Lambda$Be. Therefore,  the predicted parity reversion of
$^{12}_\Lambda$Be  suggests a possibility to probe the different deformation between the
ground and first excited states $^{11}$Be by adding an impurity of $\Lambda$ particle.

Finally, we discuss the differences between three $\Lambda N$ interactions. 
It was pointed out by Hiyama {\it et al.} \cite{Hiy97} that YNG-ND and NF have too attractive
odd-parity interaction and does not reproduce the $B_\Lambda$ of $^{9}_\Lambda$Be. They
suggested the Improved-NF  introducing a repulsive short-range part in the odd-parity interaction.
The difference between these $\Lambda N$ interactions can be found in the spectrum of
$^{12}_\Lambda$Be. We see that the $K^\pi=1/2^+\otimes \Lambda_s$ and
$3/2^-\otimes\Lambda_s$ bands are located at higher excitation energies in YNG-NF and ND
than Improved-NF. We remind the reader that the $K^\pi=1/2^-$ band of $^{11}$Be has 7
nucleons in $p$-shell, while the $K^\pi=1/2^+$ and $3/2^-$ bands have 6 and 5
nucleons. Therefore, the number of the odd-parity interactions between $\Lambda$ and
nucleons decreases for the $1/2^-_1\otimes \Lambda_s$, $1/2^+_1\otimes \Lambda_s$ and 
$3/2^-_1\otimes \Lambda_s$ bands in descending order. Consequently, when the
odd-parity interaction becomes more attractive, the energy of the $1/2^-_1\otimes \Lambda_s$ 
band is lowered relative to other two bands, or in other words, the $1/2^+_1\otimes
\Lambda_s$ and $3/2^-_1\otimes \Lambda_s$ bands are pushed up. Thus the excitation energies
of the $K^\pi=1/2^+_1\otimes \Lambda_s$ and $3/2^-_1\otimes \Lambda_s$ are sensitive to the 
odd-parity part of $\Lambda N$ interaction.

In summary, the low-lying states of $^{12}_{\Lambda}$Be have been investigated
by the HyperAMD. We predict the parity reversion of the $^{12}_{\Lambda}$Be;
the ground state parity inverted in $^{11}$Be is reverted in $^{12}_\Lambda$Be. The parity
reversion is caused by the different deformation  
of the ground and first excited states of $^{11}$Be, that produces the difference in
$B_{\Lambda}$.  This parity reversion  suggests a possibility to probe the different
deformation between the ground and first excited states $^{11}$Be by adding $\Lambda$
particle as impurity.  
 We also point out that  the excitation energies of the $K^\pi=1/2^+_1\otimes
\Lambda_s$ and $3/2^-_1\otimes \Lambda_s$ are sensitive to the  odd-parity part of the
$\Lambda N$ interaction. 

\bibliography{Parity_reversion_of_12LBe}

\end{document}